\documentclass[twocolumn,showpacs,aps,prl,superscriptaddress]{revtex4}

\usepackage{graphicx}
\usepackage{dcolumn}
\usepackage{amsmath}
\usepackage{epsfig}

\input babarsym

\newcommand{\BABARPubYear}    {05}
\newcommand{\BABARPubNumber}  {20}

\newcommand{\SLACPubNumber} {11296}

\def\figurebox#1#2#3{%
    \def\arg{#3}%
    \ifx\arg\empty
    {\hfill\vbox{\hsize#2\hrule\hbox to #2{\vrule\hfill\vbox to #1{\hsize#2\vfill}\vrule}\hrule}\hfill}%
    \else
    {\hfill\epsfbox{#3}\hfill}%
    \fi}

\begin{document}

\begin{flushleft}
\preprint{\babar-PUB-\BABARPubYear/\BABARPubNumber} 
\preprint{SLAC-PUB-\SLACPubNumber} 
\babar-PUB-\BABARPubYear/\BABARPubNumber\\
SLAC-PUB-\SLACPubNumber\\
\end{flushleft}

\title{
{\large \bf
Measurement of the $B^+\rightarrow p \bar{p} K^{+}$ Branching Fraction and Study of the Decay Dynamics
}
}

%
\author{B.~Aubert}
\author{R.~Barate}
\author{D.~Boutigny}
\author{F.~Couderc}
\author{Y.~Karyotakis}
\author{J.~P.~Lees}
\author{V.~Poireau}
\author{V.~Tisserand}
\author{A.~Zghiche}
\affiliation{Laboratoire de Physique des Particules, F-74941 Annecy-le-Vieux, France }
\author{E.~Grauges}
\affiliation{IFAE, Universitat Autonoma de Barcelona, E-08193 Bellaterra, Barcelona, Spain }
% ...
\collaboration{The \babar\ Collaboration}
\noaffiliation

\date{\today}

\begin{abstract}
With a sample of 232$\times10^{6}$ $\Upsilon(4S)\rightarrow B\bar{B}$ events 
collected with the \babar\ detector, 
we study the decay $B^{+} \rightarrow p \bar{p} K^{+}$ excluding charmonium decays to $p\bar{p}$. 
We measure a branching fraction ${\cal B}$($B^{+} \rightarrow p \bar{p} K^{+}$)=(6.7$\pm$0.5$\pm$0.4)$\times 10^{-6}$. 
An enhancement at low $p \bar p$ mass is observed and the Dalitz plot asymmetry 
suggests dominance of the penguin amplitude in this $B$ decay. 
We search for a pentaquark candidate $\Theta^{*++}$ decaying into $pK^+$ in the mass range 1.43 to 2.00$\,$GeV$/c^2$
and set limits on ${\cal B}(B^+\rightarrow \Theta^{*++}\bar{p})\times{\cal B}(\Theta^{*++}\rightarrow pK^+)$  at the $10^{-7}$ level.
 
\end{abstract}


\maketitle

This paper describes a measurement of the branching fraction of the baryonic three-body decay 
$B^+\rightarrow p\bar{p}K^+$~\cite{charge} (excluding charmonium decays to $p\bar{p}$) 
and a study of its resonant substructure. 
An earlier measurement~\cite{belle} of the branching fraction for this channel gave $(5.7^{+0.7}_{-0.6}\pm\,0.7)\times 10^{-6}$.
This channel is interesting for the dynamical information in the distribution of the three final-state particles
and for the possible presence of exotic~\cite{glue}$,\,$\cite{pent0} intermediate states. 
We also isolate decays $B^+\rightarrow X_{c\bar{c}}K^+$, where $X_{c\bar{c}}=\eta_c$ and $J/\psi$ decaying to $p\bar{p}$, 
and measure the width of the $\eta_c$.

\begin{figure}
\vspace{-0.1cm}
  \epsfig{file=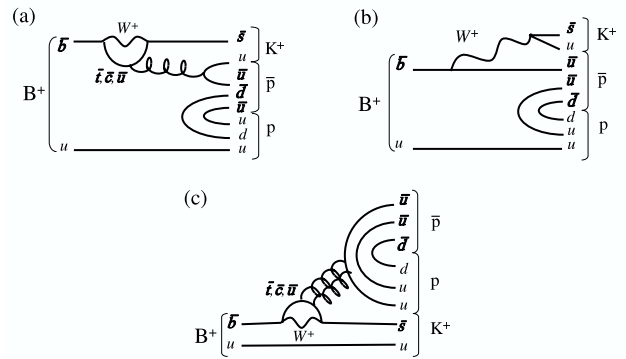,width=9cm}  
 \vspace{-2.4cm}
  \caption{
The main Feynman diagrams for the non-resonant $B^+\rightarrow p\bar{p}K^+$ decay: (a) leading penguin diagram,
(b) leading tree diagram (external $W^+$-emission), (c) Okubo-Zweig-Iizuka-suppressed penguin diagram.}  
\label{fig1}
\vspace{-0.4cm}
\end{figure}

An important feature of this decay is an enhancement at low $p\bar{p}$ masses 
reported in Ref.~\cite{belle}, similar to those that have been observed in several 
other baryonic decays of $B$~\cite{bb} and $J/\psi$~\cite{BES}. This could be a 
feature of a quasi-two-body decay in which the $p\bar{p}$ system is produced through 
an intermediate gluonic resonance~\cite{glue} (Fig.~\ref{fig1}(c)). It could also be that the decay 
is a pure three-body process and that the enhancement results from the short-range correlations between $p$
and $\bar{p}$ in the fragmentation chain~\cite{hou}$,\,$\cite{dia}.
Rosner suggested~\cite{Rosner} that it is possible to distinguish the fragmentation or 
gluonic resonance mechanisms by 
studying the distribution of events in the Dalitz plot.

The main Feynman diagrams for this decay are presented in Fig.~\ref{fig1}. 
The leading diagrams~\cite{dia} are a penguin diagram and a doubly Cabibbo-Kobayashi-Maskawa-suppressed
tree diagram shown in Fig.~\ref{fig1}(a,b). 
There is also an Okubo-Zweig-Iizuka-suppressed penguin diagram
shown in  Fig.~\ref{fig1}(c), where the $p\bar{p}$ pair is created through a pair of gluons (or a gluonic
resonance). There are four additional color-suppressed diagrams \cite{dia}: two tree diagrams 
with an internal  $W^+$-emission and a $W^+$-annihilation and two penguin 
diagrams with an internal gluon-emission that are expected to be small.
If the $p\bar{p}$ system is produced independently of the $K^+$ through a tree diagram with an external $W^+$-emission 
(Fig.~\ref{fig1}(b)) or a penguin with an external gluon-emission (Fig.~\ref{fig1}(c)),
{\it i.e.} the $p\bar{p}$ quark lines are not associated with the $\bar{s}$ or $u$
quarks in the $K^+$, then the distributions $m_{pK^+}$ and $m_{\bar{p}K^+}$ should be
identical. 
If the $u$ quark in the $K^+$ is associated with a $\bar{u}$ quark in a $\bar{p}$ 
(Fig.~\ref{fig1}(a)), larger values of $m_{pK^+}$ are favored over those of $m_{\bar{p}K^+}$~\cite{Rosner}. 
Thus a study of the Dalitz plot provides insight not only into the dominant mechanism of this decay
but also into whether the penguin or the tree amplitude is dominant.

This paper is organized as follows: first we describe the event selection and the branching-fraction measurement. 
Then we describe the $p\bar{p}$ mass spectrum and the measurement of the $\eta_c$ width.
We examine the Dalitz plot for an asymmetry between the distributions in $m_{pK^+}$ and $m_{\bar{p}K^+}$.
In the final section we describe searches for $B^+\rightarrow p\bar{\Lambda}(1520)\rightarrow p(\bar{p}K^+)$ decay  
and for the hypothesized $I=1,\,I_3=1$ pentaquark state $\Theta^{*++}$ (a member of the baryon 27-plet with quark content $uuud\bar{s}$) in the decay 
$B^+\rightarrow \Theta^{*++}\bar{p}\rightarrow (pK^+)\bar{p}$. The  
$\Theta^{*++}$ mass has been predicted~\cite{theory} to lie in the region $1.43-1.70\,\mbox{GeV/c}^2$.
 
The analysis uses 232$\times10^{6}$ $\Upsilon(4S)\rightarrow B\bar{B}$ decays
collected with the \babar\ detector~\cite{nim} at the PEP-II $e^+ e^-$ storage ring. 
Charged tracks are measured by a five-layer silicon vertex tracker (SVT) and a 40-layer drift-chamber (DCH) in a 1.5-T solenoidal magnetic field. 
A Cherenkov radiation detector (DIRC) is used for charged-particle identification.
The CsI(Tl) electromagnetic calorimeter detects photon and electron showers.
To identify kaons and protons we use $dE/dx$ measurements in
the SVT and DCH, and the pattern of Cherenkov photons in the DIRC.
The proton efficiency is 93$\%$ with 9$\%$ kaon misidentification probability.
The kaon efficiency is 87$\%$ with 2$\%$ pion misidentification probability.

We use the kinematic constraints of $B$-meson pair-production at the $\Upsilon(4S)$
to identify the $B^+\rightarrow p\bar p K^+$ signal.
Two independent variables are calculated for each $p \bar p K^+$ candidate: 
$m_{ES}=[(E^2_{cm}/2+\mbox{\bf p}_0\cdot \mbox{\bf p}_B)^2/E^2_0-{\mbox{\bf p}^2_B}]^{1/2}$ 
and $\Delta E=E^*_B-E_{cm}/2$, where $E_{cm}$ is the total center-of-mass energy,
the subscripts 0 and $B$ refer to the initial $\Upsilon(4S)$ and to the $B$ candidate, respectively, 
and the asterisk denotes the $\Upsilon(4S)$ frame. 
The resolutions on $\Delta E$ and $m_{ES}$ are about $17\,$MeV and $2.6\,$MeV/$c^2$, respectively.

Backgrounds arise primarily from random combinations in continuum events ($e^+e^-\rightarrow q\bar{q},$ where $q=u,d,s,c$). 
These  events are collimated along the original quark directions and can be distinguished from 
more spherical $B\bar{B}$ events with a Fisher discriminant~(${\cal F}$)~\cite{fisher},
a linear combination of four event-shape variables. The four variables are 
cos$\theta^*_{thr}$, the angle between the thrust axis of the reconstructed $B$ and the 
beam axis; cos$\theta^*_{mom}$, the angle between 
the momentum of the reconstructed $B$ and the beam axis; 
and the zeroth and second  Legendre polynomial momentum moments,
$L_0=\sum_{i} |{\bf p}^*_i|$ and $L_2=\sum_{i} |{\bf p}^*_i|[(3\cos^2\theta^*_{thr_{B,i}}-1)/2]$, 
where ${\bf p}^*_i$ are the momenta of the tracks 
and neutral clusters not associated with the $B$ candidate and $\theta^*_{thr_{B,i}}$ is
the angle between  ${\bf p}^*_i$ and the thrust axis of the $B$ candidate.
The event selection is optimized assuming the previously measured value of the branching fraction~\cite{belle} 
to maximize $s/\sqrt{s+b}$, where $s$ and $b$ are the expected number of 
signal and background events, respectively. The event selection retains $66\%$ of signal 
events while removing $93\%$ of continuum background.  

\begin{figure}
\vspace{-0.1cm}
\begin{center}
  \epsfig{file=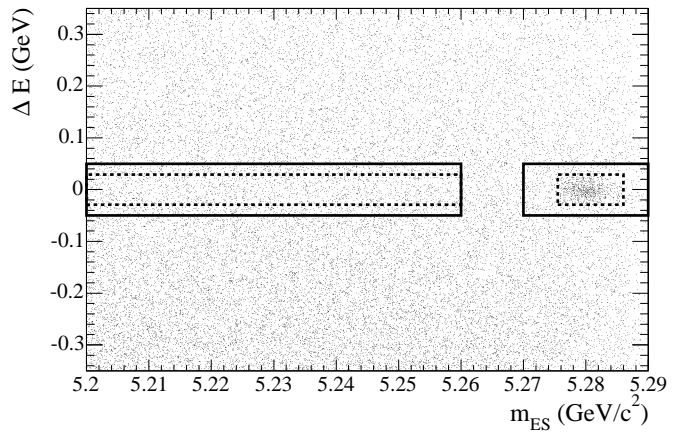,width=9cm}  
\end{center}
\vspace{-0.8cm}
  \caption{
Distribution of $\Delta E$ versus $m_{ES}$ for the $p\bar p K^+$ candidates in data.
The solid (dashed) lines define the wide (narrow) signal and sideband regions.
}  \label{fig2}
\vspace{-0.4cm}
\end{figure}

The resulting distribution of events in the $m_{ES}$$-$$\Delta E$ plane 
is shown in Fig.~\ref{fig2}. A clear signal is observed at the $B$ mass and $\Delta E$$=$0. 
Potential backgrounds are studied with Monte Carlo (MC) simulation~\cite{simu}.
The combinatorial background is expected to come predominantly ($89\%$) from continuum events. 
Background events in the signal region arise mostly from  
$B^+$$\,\rightarrow\,$$X_{c\bar{c}}(p\bar{p})K^+$, where 
$X_{c\bar{c}}$$\,=\,$$\eta_c,\,J/\psi,\,\psi',\,\chi_{c0,1,2}$ (the charmonium background), while non-charmonium 
$B$ backgrounds are expected to be negligible.
The signal and sideband regions are defined to be ``wide'' 
($5.27\,$$<\,$$m_{ES}$$<$$5.29\,$GeV/$c^2$ and $5.20$$<$$m_{ES}$$<$$5.26\,$GeV/$c^2$, $|\Delta E|$$<$$50\,$MeV) 
for the charmonium background studies 
and ``narrow'' ($5.276$$<$$m_{ES}$$<$$5.286\,$GeV/$c^2$ and $5.20$$<$$m_{ES}$$<$5.26$\,$GeV/$c^2$, $|\Delta E|$$<$$29\,$MeV)
for the Dalitz plot study.
 
\begin{figure}
\begin{center}
\epsfig{file=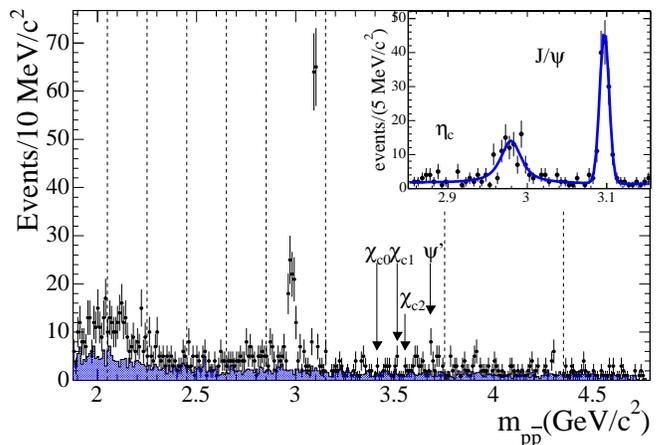,width=9cm}  
\end{center}
\vspace{-0.8cm}
  \caption{The $m_{p\bar{p}}$ distribution for data in the wide signal (points) and sideband (shade)
regions. The sideband histogram is scaled to the expected number of the combinatorial background events in the signal region.}
\label{fig3}
\vspace{-0.4cm}
\end{figure}

To extract the $p \bar{p}K^+ $ signal yield, we fit the $\Delta E$ distributions for candidates that lie in the
5.27$<$$m_{ES}$$<$5.29$\,$GeV/$c^2$ region separately in nine bins of $m_{p\bar{p}}$. 
The size of the bins is shown in Fig.~\ref{fig3}. 
We use a linear function for the background and a double Gaussian distribution for the signal. 
The widths and means of the Gaussian distributions and their relative areas are fixed to values obtained from MC simulation, which 
is also used to calculate the detection efficiency ($\eps_{m_{p\bar p}}$) 
in each $m_{p\bar{p}}$ bin. Across the allowed kinematic region,
$\eps_{m_{p\bar p}}$ declines smoothly from $30\%$ at threshold to 
$24\%$ at the highest mass. The $\Delta E$ fits for $m_{p\bar{p}}$ below $2.85\,$GeV/$c^2$ 
yield 343$^{+27}_{-26}$ signal events. From the known number of charged  
$B$ mesons in the sample, the branching fraction for $m_{p\bar{p}}$ below the $\eta_c$ mass is measured to be 
${\cal B}(B^+\rightarrow p\bar{p}K^+;\,m_{p\bar{p}}$$<$$2.85\,$GeV/$c^2)=(5.3\pm 0.4\pm 0.3)\times 10^{-6}$. 
 
An estimate of the number of 
charmonium events in the $m_{p\bar{p}}$$>$$2.85\,$GeV/$c^2$ region is required
to determine the total non-charmonium branching fraction. 
To minimize the systematic error on that quantity, we fit the $m_{p\bar{p}}$
spectrum for the number of the non-charmonium events in the 
primary ``charmonium'' region (2.85$<$$m_{p\bar{p}}$$<$$3.15\,$GeV/$c^2$). 
To improve the ${p\bar{p}}$ mass resolution in the $m_{p\bar{p}}$ fit, 
we perform a kinematic fit fixing the mass and energy of each
$B$ candidate in the wide signal and sideband regions to their known values.
The $m_{p\bar{p}}$ distribution is shown in Fig.~\ref{fig3}, where prominent
signals for the $\eta_c$ and $J/\psi$ decaying into $p\bar{p}$ are visible.   
The region used in the $m_{p\bar{p}}$ fit, 2.4$<$$m_{p\bar{p}}$$<$$3.4\,$GeV/$c^2$, is chosen wider than
the ``charmonium'' region of interest (2.85$<$$m_{p\bar{p}}$$<$$3.15\,$GeV/$c^2$), shown in Fig.~\ref{fig3}(inset),
to improve the statistical uncertainties on the $p\bar{p}K^+$ signal and combinatorial background yield. 
The $\eta_c$ peak is described by a convolution of a Breit-Wigner distribution and a Gaussian distribution, and 
the $J/\psi$ peak by a sum of two Gaussian distributions with a common mean. The shapes are obtained from MC simulation.
The width of the broader $J/\psi$ Gaussian distribution and ratio of areas of the two  $J/\psi$ Gaussian distributions are constrained in the fit to 
their MC values. A common width is used for the narrow Gaussian distributions for $J/\psi$ and $\eta_c$ 
and is a free parameter in the fit. 
The $p\bar{p}K^+$ signal and combinatorial background distributions are modeled by 
a linear function of $m_{p\bar{p}}$. 
The inset of Fig.~\ref{fig3} shows this fit, which results in 114$^{+15}_{-14}$
$\eta_c $ events and 137$^{+13}_{-12}$ $J/\psi$ events. Correcting for the detection
efficiency of $(26.9\pm0.2)\%$, we find 
${\cal B}(B^+\rightarrow \eta_cK^+)\times{\cal B}(\eta_c\rightarrow p\bar{p})=(1.8^{+0.3}_{-0.2}\pm 0.2)\times 10^{-6}$ and
${\cal B}(B^+\rightarrow J/\psi K^+)\times{\cal B}(J/\psi\rightarrow p\bar{p})=(2.2\pm{0.2}\pm 0.1)\times 10^{-6}$ 
in agreement with the accepted values~\cite{PDG}. The fit yields a total $\eta_c$ width of $\Gamma(\eta_c)$$=$$25^{+6}_{-5}$$\pm3\,$MeV/$c^2$ 
consistent with the current values~\cite{PDG} and a mass resolution of $5.7\pm0.4\,$MeV/$c^2$ in agreement with MC expectations.

\begin{figure}
\begin{center}
\epsfig{file=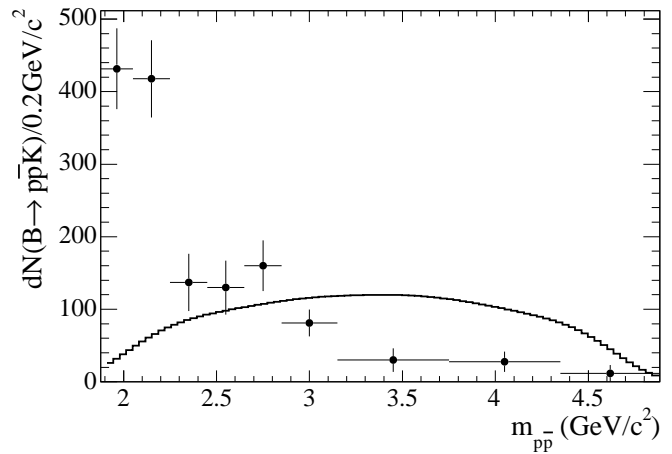,width=9cm}  
\end{center}
\vspace{-0.6cm}
  \caption{Efficiency-corrected yield of $B^+\rightarrow p\bar{p}K^+$ events as a function of $m_{p\bar{p}}$
in data (points) and in three-body phase-space signal MC (histogram). Errors shown are statistical only.}
\label{fig4}
\vspace{-0.2cm}
\end{figure}

The $m_{p\bar{p}}$ fit yields 88$\pm$6 $p \bar{p}K^+ $ signal and 
combinatorial background events in the charmonium 
region (see Fig.~\ref{fig3}).
In this region, the latter contribution is estimated from the $\Delta E$ fit to be 53$\pm$5 events, resulting in a 
non-charmonium $p \bar{p}K^+ $ signal of 35$\pm$8 events. 
The contribution from higher-mass charmonium modes is estimated 
to be 24$\pm$5 events from the accepted~\cite{PDG} values for their branching fractions.
Adding the $p \bar{p}K^+ $ signal yield obtained from the $\Delta E$ fits outside the 
``charmonium'' region with non-charmonium $p \bar{p}K^+ $ signal in the ``charmonium'' region, 
and subtracting the contribution of the higher mass charmonium modes results in a total non-charmonium signal yield of 433$\pm$33 events.
Correcting the signal yield for efficiency in each of the $m_{p\bar{p}}$ bins 
and normalizing to the number of $B^+$ mesons in the data sample results in a total branching fraction of 
${\cal B}(B^+\rightarrow p\bar{p}K^+)=(6.7\pm 0.5\pm 0.4)\times 10^{-6}$ with charmonium decays to $p\bar{p}$ excluded. 
Figure~\ref{fig4} shows the background-subtracted and efficiency-corrected 
$p\bar{p}$ mass spectrum and the expectation for a three-body phase-space decay. 
The existence of a low-mass enhancement in the $p\bar p$ mass as previously observed by Belle~\cite{belle} is confirmed. 

The charge asymmetry is defined as $A_{ch}$$=$$(N_{B^-}$$-$$N_{B^+})$$/(N_{B^-}$$+$$N_{B^+})$, 
where $N_{B^{\pm}}$ is the number of $B^{\pm}$$\rightarrow$$p\bar{p}K^{\pm}$ events. 
We use the same fitting procedure as for the branching fraction measurement, 
and find $A_{ch}$$=$$-0.16^{+0.07}_{-0.08}\pm0.04$ for 
$m_{p\bar{p}}$$<$$2.85\,$GeV/$c^2$.

\begin{figure}
\begin{center}
\epsfig{file=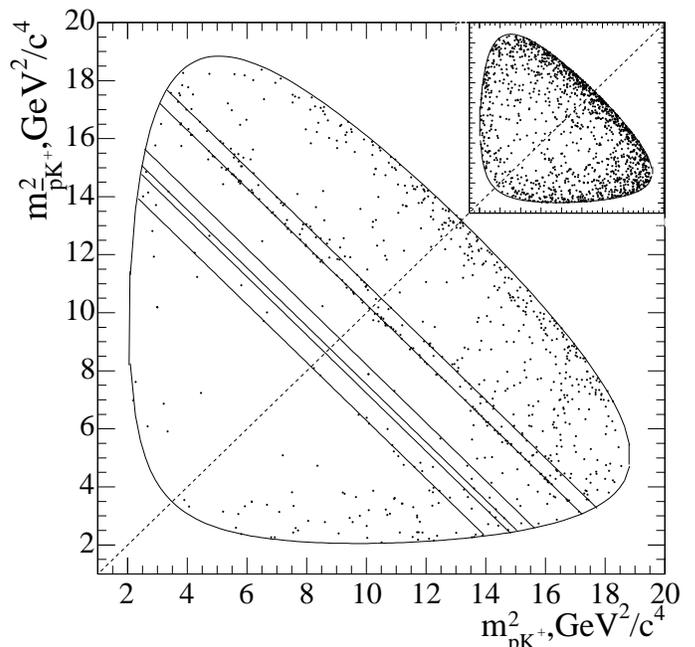,width=9cm}  
\end{center}
\vspace{-.7cm}
  \caption{
Dalitz plot of data in the narrow signal region, and  sideband region (inset). These distributions are not efficiency-corrected.
The lines show the positions of the prominent charmonium backgrounds, from left to right $\psi',\,\chi_{c2,1,0},\,J/\psi,\,\eta_c$. 
The sideband contains about eight times more combinatorial events than are expected in the signal region.}
\label{fig5}
\vspace{-0.4cm}
\end{figure}

For the remainder of this paper to increase the signal purity, only events in the narrow signal and $m_{ES}$-sideband regions
are considered. After selecting the $B$ candidates, 
we perform a kinematic fit for each $B$ candidate, 
fixing its mass and energy to their known values. 

We study the dynamics of the three-body decay by constructing signal and sideband Dalitz plots (Fig.~\ref{fig5}).  
There are 780 (1661) events in the signal (sideband) region. The sideband contains about eight times
more combinatorial events than the signal region. 
The Dalitz plot for the signal region shows the threshold enhancement in the $p\bar{p}$ mass spectrum, 
as well as clear diagonal bands corresponding to $\eta_c$, $J/\psi$ and $\psi'$ decays. 

\begin{figure}
\begin{center}
\epsfig{file=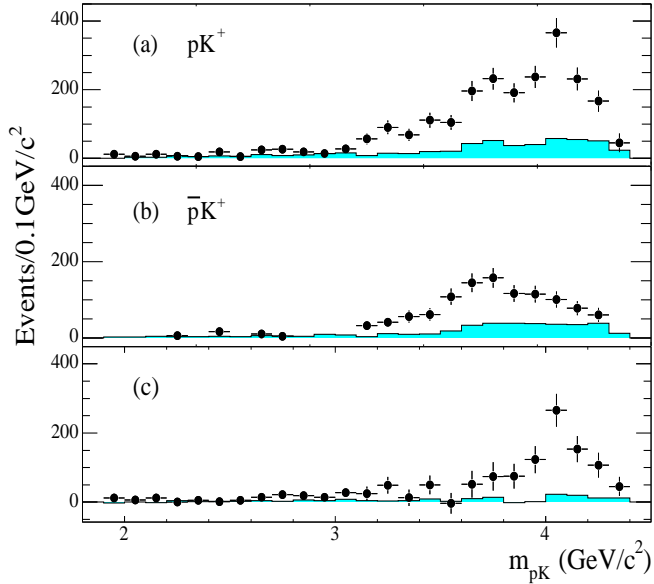,width=9cm,height=8cm}  
\end{center}
\vspace{-0.7cm}
  \caption{ Efficiency-corrected distributions in the narrow signal (points) and rescaled sideband (shade) regions:
(a) $m_{pK^+}$ (for $m_{pK^+}>m_{\bar{p}K^+}$), (b) $m_{\bar{p}K^+}$ (for $m_{pK^+}<m_{\bar{p}K^+}$), and
(c) difference between (a) and (b). Errors shown are statistical only. }
\label{fig6}
\vspace{-0.4cm}
\end{figure}

To study the $m_{pK^+}$ and $m_{{\bar p}K^+}$ asymmetry, we divide the Dalitz plot 
along the $m_{pK^+}=m_{{\bar p}K^+}$ line (dashed line in Fig.~\ref{fig5}) 
and each of the two halves is projected onto the nearer axis. 
The corresponding distributions for the events in signal and rescaled sideband regions are 
shown in Fig.~\ref{fig6}(a,b). 
No asymmetry is expected from variations in $\eps_{m_{p\bar p}}$ which is charge-symmetric 
and slowly varying with ${m_{p \bar p}}$, nor from the small combinatorial background shown in Fig.~\ref{fig6}(a,b). 
The asymmetry appears as a broad enhancement peaking at about 4$\,$GeV in the $pK^+$ combinations (Fig.~\ref{fig6}(c)).
This could be an indication of a correlation 
between quarks in $\bar{p}$ and $K^+$ if the $B$ decay proceeds through a penguin
diagram (Fig.~\ref{fig1}(a)). No quantitative theoretical description of
this correlation is available at the moment.

\begin{figure}
\vspace{-0.3cm}
\begin{center}
\epsfig{file=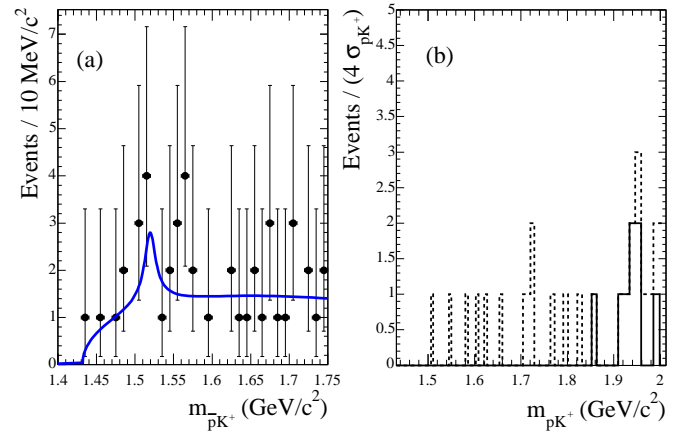,width=9cm}  
\end{center}
\vspace{-0.8cm}
  \caption{(a) The $m_{\bar{p}K^+}$ distribution for data events in $\Lambda(1520)$
mass region; (b) The $m_{pK^+}$ distributions for data events in the signal region outside (dashed) or inside (solid) 
the 2.85$<$$m_{p\bar{p}}$$<$3.15$\,$GeV/$c^2$ region. 
Distributions are not efficiency corrected.}
\label{fig7}
\vspace{-0.5cm}
\end{figure}

The two-body decay $B^+\rightarrow p\bar{\Lambda}(1520)$ could also be present in the $p \bar{p}K^+ $ signal sample.
The efficiency for detection of this channel is determined in dedicated MC simulation to be $(4.7\pm0.1)\%$,
including ${\cal B}(\Lambda(1520)\rightarrow pK^-$)~\cite{PDG}. 
The $m_{\bar{p}K^+}$ spectrum, shown in Fig.~\ref{fig7}(a), is  fit with an ARGUS function~\cite{Argus} 
for the background and a Breit-Wigner convolved with a double Gaussian 
(with a common mean) for the $\Lambda(1520)$ signal shape. The mass resolutions and the ratio of areas of the Gaussians
are fixed to the values obtained from MC simulation, while we fix the mean and the natural 
width to established values~\cite{PDG}; the endpoint of the ARGUS function is 
fixed to the sum of the proton and kaon masses. An unbinned maximum likelihood fit (Fig.~\ref{fig7}(a)) 
results in an upper limit (U.L.) on ${\cal B}(B^+\rightarrow p\bar{\Lambda}(1520))$ of $1.5\times 10^{-6}$ at $90\%$ C.L.
 (including a systematic error of 16$\%$).

The search for a light $\Theta^{*++}$ pentaquark candidate ($m_{\Theta^{*++}}$$<$2$\,$GeV/$c^2$)~\cite{pent} 
proceeds as follows. 
From $B^+\rightarrow p\bar{p}K^+$ three-body phase-space MC as well as five dedicated
signal MC samples with $m_{\Theta^{*++}}$$\,=\,$$1.5,\,1.6,\,1.7,\,1.8,\,1.9\,$GeV/$c^2$,
we find the mass resolution ($\sigma_{pK^+}$) to vary from 1.0 to 3.0$\,$MeV/$c^2$ for 
1.43$<$$m_{pK^+}$$<$2.00$\,$GeV/$c^2$, 
and the average reconstruction efficiency to be $(20.5\pm 0.1)\%$ in this mass region.
The events with $m_{p\bar{p}}$ in the charmonium region are vetoed.
The $pK^+$ mass distribution of the remaining events is shown in Fig.~\ref{fig7}(b). 
A Bayesian approach is used to calculate the U.L. at 90$\%$ C.L. as a function of $m_{pK^+}$, 
assuming Poisson-distributed events in the absence of background.  
Each limit is increased by the total systematic error of $6\%$. The U.L. for 
${\cal B}(B^+\rightarrow \Theta^{*++}\bar{p})\times{\cal B}(\Theta^{*++}\rightarrow pK^+)$ 
is measured to be $0.5\times 10^{-7}$ for $1.43<m(\Theta^{*++})<1.50\,\mbox{GeV/c}^2$,
$<0.9\times 10^{-7}$ for $1.50<m(\Theta^{*++})<1.72\,\mbox{GeV/c}^2$, and
$<1.2\times 10^{-7}$ for $1.72<m(\Theta^{*++})<2.00\,\mbox{GeV/c}^2$.

\begin{table}\caption{Systematic uncertainties in percent on the branching fraction measurements 
and in the values of uncertainties for the symmetry measurements. Values for $m_{p\bar{p}}$ below $2.85\,$GeV/$c^2$
are given in brackets.}
\label{systematics}
\begin{tabular}{|l|c|c|c|c|c|}
\hline
{Type}&{$p\bar{p}K^+$}&{$\eta_c K^+$}&{$p\bar{\Lambda}(1520)$}&{$\bar{p}\Theta^{*++}$}&{$A_{ch}$}\\\hline
{$B$-counting}          &{1.1(1.1)}&{1.1}&{1.1}&{1.1}&{$-$}\\
{Tracking/PID}              &{3.8(3.8)}&{3.4}&{4.2}&{4.2}&{0.02}\\
{MC Statistics}&{2.1(2.4)}&{0.7}&{1.0}&{0.5}&{0.03}\\
{B.F. Errors}        &{$0.9(-)$}&{$-$}&{2.2}&{$-$}&{$-$}\\
{Selection}        &{$0.2(-)$}&{0.4}&{3.9}&{3.9}&{$-$}\\
{$\Delta E$/Mass Fits}     &{3.6(2.4)}&{8.9}&{14.3}&{$-$}&{0.01}\\\hline
{Total}                 &{5.8(5.2)}&{13.5}&{15.6}&{6.1}&{0.03}\\\hline
\end{tabular}
\end{table}

The systematic uncertainties for each analysis are summarized in Table \ref{systematics}.
The $\Upsilon(4S)$ is assumed to decay equally to $B^0\bar{B^0}$ and $B^+B^-$ mesons. 
Incomplete knowledge of the luminosity and cross-section leads to a $1.1\%$ uncertainty. 
Charged-tracking and particle-identification (PID) studies in the data lead to small corrections applied to each track in these simulations. 
Limitation of statistics and purity in these data-MC comparisons lead to 
residual tracking/PID uncertainties. A large control sample of
$B^+$$\rightarrow$$J/\psi(e^+e^-)K^+$ is separately 
studied in data and MC simulations to understand the residual errors from the event-shape, $\Delta E$, and $m_{ES}$ cuts. 
Limitation of MC statistics employed in each analysis contributes to a small uncertainty. 
Branching fraction uncertainties (B.F. Errors) \cite{PDG} on 
${\cal B}(B^+\rightarrow XK^+)\times{\cal B}(X\rightarrow p\bar{p})$, where $X=\chi_{c[0,1,2]},\psi'$ 
and ${\cal B}(\Lambda(1520)\rightarrow pK^-)$ affect the 
total $p{\bar p}K^+$ and the $p\bar{\Lambda}$ branching fraction measurements, respectively. 
Where the MC values are used to fix signal shape parameters in a fit, the parameters are varied within their 
uncertainties and the data are refit to propagate this uncertainty.  In a similar fashion, different
ranges and background functions are employed to establish the uncertainty on the mass spectra fits 
(resulting, for example, in the $\Gamma(\eta_c)$ uncertainty of 3$\,$MeV). 

In summary, with 210$\,\mbox{fb}^{-1}$ of data, 
we isolate the $B^+\rightarrow p\bar{p}K^+$ final state, and measure its 
non-charmonium branching fraction to be 
$(5.3\pm 0.4\pm 0.3)\times 10^{-6}$ for $m_{p\bar{p}}$ below $2.85\,$GeV/$c^2$ and 
$(6.7\pm 0.5\pm 0.4)\times 10^{-6}$ for the whole $m_{p\bar{p}}$ range. 
We measure $A_{ch}$$=$$-0.16^{+0.07}_{-0.08}\pm0.04$
for $m_{p\bar{p}}$ below $2.85\,$GeV/$c^2$.
The existence of a low-mass enhancement of the $p\bar p$ pair is confirmed. 
The asymmetry between $pK^+$ and ${\bar p}K^+$ final states in the Dalitz plot is demonstrated, 
providing evidence supporting the dominance of the penguin amplitude in this $B$ decay.  
We measure the total width of $\eta_c$ to be $25^{+6}_{-5}$$\pm3\,$MeV/$c^2$. 
An upper limit of the decay rate to $p\bar{\Lambda}(1520)$ is set at $1.5\times 10^{-6}$. 
No evidence is found for the pentaquark candidate $\Theta^{*++}$
in the mass range 1.43 to 2.0$\,$GeV/$c^2$, decaying into $pK^+$, 
and branching fraction limits are established at the $10^{-7}$ 
level.

We thank S.J.~Brodsky for useful discussions.
The collaboration is grateful for the excellent luminosity and machine conditions
provided by our \pep2\ colleagues, 
and for the substantial dedicated effort from
the computing organizations that support \babar.
The collaborating institutions wish to thank 
SLAC for its support and kind hospitality. 
This work is supported by
DOE
and NSF (USA),
NSERC (Canada),
IHEP (China),
CEA and
CNRS-IN2P3
(France),
BMBF and DFG
(Germany),
INFN (Italy),
FOM (The Netherlands),
NFR (Norway),
MIST (Russia), and
PPARC (United Kingdom). 
Individuals have received support from CONACyT (Mexico), A.~P.~Sloan Foundation, 
Research Corporation,
and Alexander von Humboldt Foundation.


\begin{thebibliography}{99}
\bibitem{charge} Charge-conjugate reactions are included implicitly throughout the paper.
\bibitem{pent} The $\Theta^{*++}$ width is assumed to be negligible compared to $pK^+$ mass resolution.
\end{thebibliography}
\end{document}